# High-resolution ARPES endstation for *in-situ* electronic structure investigations at SSRF

Yi-Chen Yang [1,2] Zheng-Tai Liu [1,2,*] Ji-Shan Liu [1,2] Zhong-Hao Liu [1,2]
Wan-Ling Liu [1,2] Xiang-Le Lu [1,2] Hong-Ping Mei [1,2] Ang Li [1,2]
Mao Ye [1,2] Shan Qiao [1,2] Da-Wei Shen [1,2,**]

**Affiliations:**

[1]State Key Laboratory of Functional Materials for Informatics, Shanghai Institute of Microsystem and Information Technology, Chinese Academy of Sciences, Shanghai 200050, China

[2]University of Chinese Academy of Sciences, Beijing 100049, China

[*]Corresponding author. *E-mail address:* ztliu@mail.sim.ac.cn

[**]Corresponding author. *E-mail address:* dwshen@mail.sim.ac.cn

**Abstract:** Angle-resolved photoemission spectroscopy (ARPES) is one of the most powerful experimental techniques in condensed matter physics. Synchrotron ARPES, which uses photons with high flux and continuously tunable energy, has become particularly important. However, an excellent synchrotron ARPES system must have features such as a small beam spot, super-high energy resolution, and a user-friendly operation interface. A synchrotron beamline and an endstation (BL03U) were designed and constructed at the Shanghai Synchrotron Radiation Facility. The beam spot size at the sample position is 7.5 (V) μm × 67 (H) μm, and the fundamental photon range is 7–165 eV; the ARPES system enables photoemission with an energy resolution of 2.67 meV@21.2 eV. In addition, the ARPES system of this endstation is equipped with a six-axis cryogenic sample manipulator (the lowest temperature is 7 K) and is integrated with an oxide molecular beam epitaxy system and a scanning tunneling microscope, which can provide an advanced platform for *in-situ* characterization of the fine electronic structure of condensed matter.

**Keywords:**   Synchrotron·ARPES·*in-situ*·VUV laser·quantum materials

## 1 Introduction

The establishment of band theory is a milestone in condensed matter physics, for it describes the mechanism of conductivity and provides theoretical guidance for the rapid development of the semiconductor industry. The band theory of solids involves a one-electron approximation in which an electron moves in a periodic potential representing the nucleus and the averaged potential of other electrons [1]. However, some unexpected discoveries demonstrated the shortcomings of this theory. For instance, NiO, MnO, and CoO, which were predicted to be metals by band theory, were eventually found to be insulators. Further studies revealed that electron–electron or electron–phonon interactions in these materials are nonnegligible, and therefore the one-electron approximation does not apply to these many-body systems [2]. These phenomena are attributed to so-called strongly correlated physics, including high-temperature superconductivity, the Mott metal–insulator phase transition, and other fascinating frontiers of condensed matter physics [3–5].

In this context, the prediction of the electronic properties of such strongly correlated electron systems using band theory is far more challenging, in contrast to the case of conventional semiconductors, which can be well described by one-electron theories. However, angle-resolved photoemission spectroscopy (ARPES) has been proven to be a reliable tool to experimentally probe the fine electronic properties of these correlated materials [6–8]. Specifically, this technique can provide straightforward information on the momentum-resolved low-energy electronic structure of matter in the vicinity of the Fermi level, which is vital to detecting various properties of solids. Therefore, in the past few decades, ARPES has become one of the most popular and widely used experimental techniques for studying the electronic structure of correlated electron systems [9–11].

However, the recent development of quantum materials poses new challenges for this technique. Conventional ARPES setups utilize gas discharge lamps, whose radiation energy is dependent on the excited gas but is unpolarized, as light sources. The limitations on the photon energy, flux, and beam spot size of gas discharge lamps seriously hinder new discoveries in condensed matter physics, in particular for fine gap structure in novel unconventional superconductors and complex surface/bulk states in topologically nontrivial materials; thus, it is necessary to develop advanced spectroscopic probes. In particular, owing to the limited photon energy, the detection of the three-dimensional dispersion in momentum space using conventional ARPES is limited. The construction of third-generation synchrotron facilities offers windows of opportunity for the development of ARPES. Photons are produced by passing accelerated electrons through a undulator, and then a monochromator is used to obtain an intense light flux with a bandwidth down to the meV scale. In addition, the photon energy of the synchrotron radiation beam can easily be tuned continuously, which makes it possible for ARPES to disentangle the surface and bulk states of solids. ARPES beamlines with tunable synchrotron radiation have been installed worldwide [12,13]. Some are designed for the vacuum ultraviolet (VUV) region with high energy resolution [14–16]. By using a high flux and small beam spot, the efficiency and momentum resolution of ARPES can be improved greatly, and even sub-meV and microscale measurements can be realized [17,18]. Moreover, the polarization of the radiation is adjustable; therefore, researchers can utilize matrix element effects to distinguish bands with different orbital characters.

Another serious difficulty in ARPES measurement is the need to obtain clean and flat surfaces of solids samples owing to the extreme surface sensitivity of this technique. For conventional ARPES measurements, these clean surfaces can be obtained by cleaving layer-structured samples or through sputtering-annealing cycles. However, for the increasingly fascinating quantum materials, such as three-dimensional uncleavable solids and correlated interfaces/heterostructures, it is challenging to obtain a flat surface along the desired crystalline orientation by cleavage. In addition, sputtering-annealing cycles are not suitable for most compounds because the process may change the surface structure and composition. To overcome these difficulties, epitaxial growth techniques are applied to realize *in-situ* ARPES measurements. By connecting a growth chamber to the ARPES setup and transferring freshly grown samples under ultrahigh vacuum (UHV), clean and atomically flat surfaces can be prepared for ARPES investigation. In the past, most *in-situ* photoemission measurements have been realized by integrating pulsed laser deposition and ARPES, and the recent integration of molecular beam epitaxy (MBE) with ARPES further advanced the *in-situ* synthesis-probe approach [19,20]. Owing to the relatively low kinetic energies of atomic beams and the low partial pressure of oxidants, higher crystal quality and sharper interfaces superior to those produced by other thin film synthesis methods can be obtained. In addition, atomic layer-by-layer MBE growth enables good control of the desired termination, facilitating the probing of specific surfaces of complex crystalline samples. Moreover, this technique can be used to fabricate metastable phases and even heterostructures and superlattices of a single crystalline phase, which effectively expands the scope of the investigation of novel quantum materials. Nevertheless, intentionally designed ARPES setups based on synchrotron radiation beamlines, which are capable of high-resolution *in-situ* measurements, are still rare worldwide.

To meet new challenges to conventional ARPES, we designed and constructed a specialized ARPES beamline, BL03U, at the Shanghai Synchrotron Radiation Facility (SSRF). This endstation is integrated with a high-resolution ARPES instrument, an ultralow-temperature scanning tunneling microscope, and an oxide MBE (OMBE) setup, which can realize epitaxial film growth and high-resolution *in-situ* electronic structure measurements in both momentum and real space. Furthermore, a VUV laser light source with a photon energy range of 5.9–7 eV is utilized in this endstation to improve the ARPES performance. This endstation is open to users and has enabled the discovery of many important materials, such as the bilayer superconductor Fe–As [21], magnetic Weyl semimetal $Co_3Sn_2S_2$ [22], magnetic topological insulator $MnBi_2Te_4$ [23–25], and topological nodal-line semimetal $SrAs_3$ [26].

# 2 Light Source

Synchrotron ARPES is most commonly operated at photon energies of 20–100 eV, and the photoelectron mean free path in this energy range is at the universal minimum in solids, as in the gas discharge sources used for laboratory-based ARPES. Thus, for comprehensive study of electronic structure in solids, including surface and bulk states, a synchrotron radiation beamline that can cover low photon energies (down to 7 eV) and a VUV laser system were constructed in our ARPES endstation.

## 2.1 Beamline layout

The BL03U beamline at the SSRF is designed for high-resolution ARPES. To generate low-energy photons (7–165 eV) and solve the heat load problem at the high-energy storage ring (3.5 GeV), a specially designed APPLE-Knot undulator is adopted in this beamline [27]. In addition to the fundamental photon energy of the APPLE-Knot undulator, which covers 7–165 eV, higher-energy photons can be obtained from higher-order harmonic components. The optical layout of the beamline is shown in Fig. 1 [28]. The photon beam emitted from the undulator is deflected and focused by the first cylindrical mirror (M1), and apertures AP1 and AP2, before and after M1, can control the acceptance angle of the photon beam. The plane mirror M2 reflects the photon beam upward to the varied-line-space plane grating (GR), and the beam is focused vertically onto the exit slit by GR. In addition, the plane mirror M5 can switch the photon beam to pass the main or branch lines. After the exit slit, an ellipsoidal mirror (M3) focuses the photon beam again in both the horizontal and vertical directions toward the sample position in the ARPES chamber. To obtain photons with a wider energy range, the monochromator is equipped with three optical gratings: G1, 188 l/mm; G2, 620 l/mm; and G3, 2000 l/mm, which cover a photon energy range of 7–798 eV.

The energy resolution of the monochromator was tested by gas-phase photoabsorption measurements of He and Ne. When the G2 grating (620 l/mm) was used, the fitting of the experimental data gave an energy resolution of 0.47 meV at 21.6 eV, which corresponds to an energy resolving power of up to 45,000. In addition, the energy resolution was 1.03 and 2.11 meV at 48.4 and 65.5 eV, respectively. The photon flux was measured using a photodiode downstream of M3, which was calibrated using a standard curve from the diode manufacturer, Opto Diode Corporation. For a 240 mA ring current and 30 μm vertical width of the exit slit, the photon flux remains above $1.6 \times 10^{13}$ photons/s/0.1%B.W. (0.1% bandwidth) in an energy range of 21–65 eV for both G1 and G2. Moreover, the spot size of the focused photon beam at the ARPES sample position has been experimentally measured by a knife-edge device. When the vertical width of the exit slit is set to 30 μm at a photon energy of 21.6 eV, the beam spot is as small as 7.5 μm × 67 μm (detailed data are presented in sec. 4.1), which ensures the optimal angular resolution in ARPES measurements.

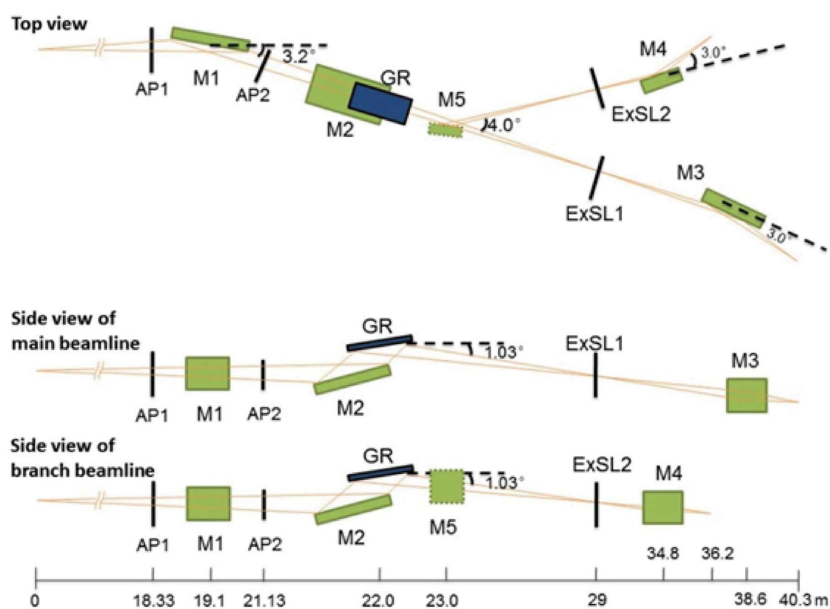

**Fig. 1.** Schematic layout of BL03U beamline. Adapted from **Ref. 28** with permission.

## 2.2 Laser system

The VUV laser is a desirable and powerful light source for ARPES measurements; it has many advantages, such as an intense photon flux, a very small spot size, and high momentum and energy resolution [29,30]. The photon energy range of VUV lasers is typically 6–11 eV, and the corresponding mean free path of photoelectrons is larger than that of conventional discharge lamps, which is crucial for probing the bulk state of solids. With the development of near-ultraviolet solid-state sources, in particular the techniques of second harmonic generation (SHG) and high harmonic generation (HHG), the VUV laser has become a popular light source for state-of-the-art ARPES [31,32]. For example, the HHG technique has been used to generate 11 fs pulses for time-resolved ARPES [33], and Mao et al. [34] obtained the smallest (~0.76 μm) focal spot of a VUV laser, 177 nm, by using a flat lens without spherical aberration, which can be utilized in spatially resolved ARPES. On the basis of these considerations, a VUV laser light source was constructed at the BL03U endstation to complement the synchrotron radiation and thus improve the ARPES performance. The VUV laser is produced by SHG processes in a nonlinear optical crystal. A tunable Ti:sapphire picosecond laser oscillator (Spectra-Physics Tsunami) is used as a seed source; it utilizes a 532 nm green laser as the pump source. Its wavelength can be tuned in the infrared band from 700 to 850 nm. The nonlinear optical crystal plays a key role in the SHG processes. Considering the performance of different crystals at different wavelengths, a cascade of two frequency-doubling stages in nonlinear $BaB_2O_4$ (BBO) crystal and $KBe_2BO_3F_2$ (KBBF) crystals is utilized. The infrared beam is up-converted to the second harmonic using the BBO crystal. Then the SHG ultraviolet beam is introduced into the $N_2$-filled frequency-doubling chamber and focused onto the KBBF crystal through a quartz window by a reflection mirror. Next, a KBBF prism-coupled device [35] is employed to generate deep-ultraviolet (DUV) coherent light by SHG, whose photon energy can be tuned from 5.9 to 7 eV. Fig. 2 displays a schematic diagram of the typical 6.89 eV laser generation processes. The DUV beam is focused by an antireflection-coated $CaF_2$ lens onto the sample position in the ARPES analysis chamber, and the beam spot at the sample position is found to be less than 30 μm × 30 μm.

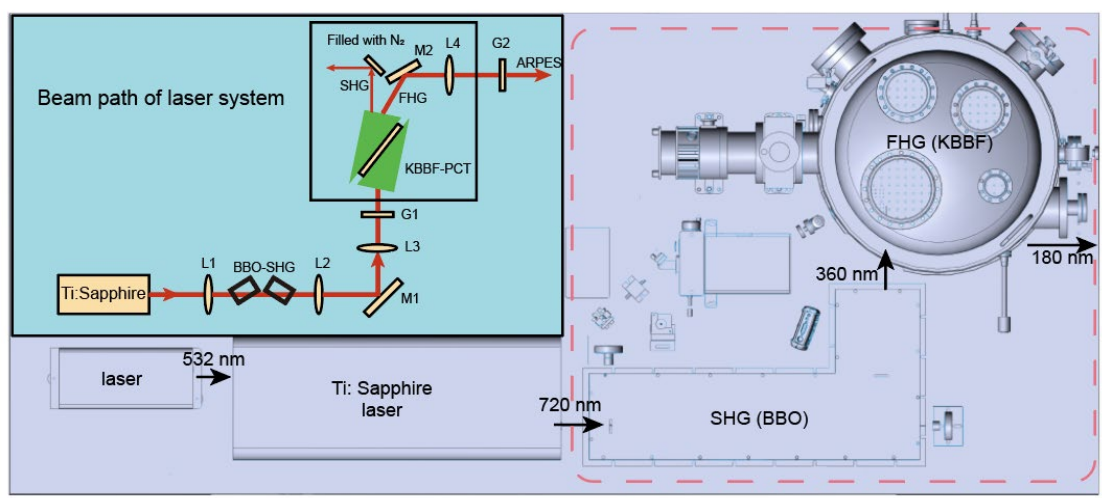

**Fig. 2.** Schematic layout of VUV laser optical system. Inset shows main beam path.

# 3 Endstation Design

## 3.1 Overview

As shown in the schematic diagram (Fig. 3), the BL03U endstation consists of five main parts: a load-lock chamber, an OMBE system, an ARPES system, a scanning tunneling microscopy (STM) system, and a UFO-like transfer chamber. The analysis chamber (lower chamber) of the ARPES system is connected to two symmetrical incident light sources (the synchrotron radiation and the VUV laser), which can be conveniently switched. The UFO chamber serves as a transfer station and is connected with all the other main functional chambers to implement sample transfer between different chambers under UHV and ensure fresh and clean sample surfaces for further photoemission experiments. A specially designed transfer arm in the UFO chamber is used to grab/place samples. The rotation of the transfer arm is controlled by a programing motor, which can move samples to suitable positions accurately. In addition, the ARPES chamber is equipped with auxiliary devices, including a low-energy electron diffraction (LEED) system and three alkali metal evaporators. This endstation can satisfy the requirements for measurements of most types of single-crystal and thin film samples. However, the sample should be a conductor or small-gap semiconductor to avoid charging effects during ARPES measurements. This endstation also has auxiliary equipment that can provide pretreatments such as sputtering-annealing processes/surface doping and *in-situ* LEED measurements.

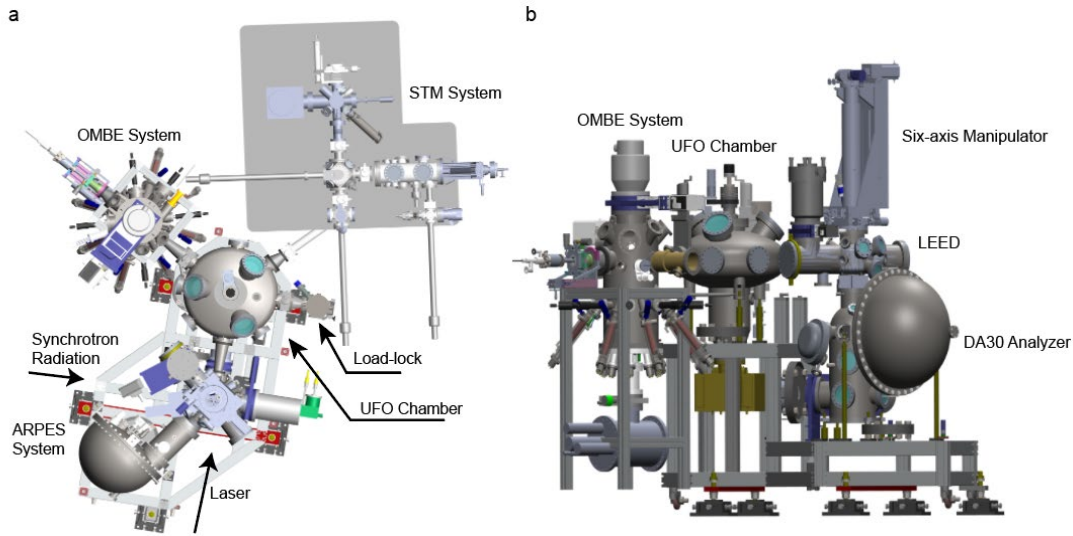

**Fig. 3.** Schematic of BL03U endstation. **a** Top view of endstation. **b** Side view of ARPES-OMBE system.

### 3.2 Angle-resolved photoemission spectroscopy

The ultrahigh-resolution ARPES system consists of four main parts: a pre-chamber, an analysis chamber, a low-temperature sample manipulator, and a DA30 electron analyzer. The detailed design of this system is shown in Fig. 4. For ARPES measurements, the data quality depends strongly on the sample surfaces. To avoid surface contamination and the scattering of the emitted electrons, the system must be maintained in a UHV environment. Therefore, multiple pumps, including turbomolecular pumps, ion pumps, titanium sublimation pumps, and getter pumps, are typically used to keep the vacuum pressure below 1E-10 mbar. A state-of-the-art six-axis open-cycle cryogenic manipulator that is suitable for a UHV and low-temperature environment is integrated to control the sample translational motion (XYZ) and orientation (polar/tilt/azimuthal). In this design, continuous 360° rotation around the polar axis is possible. The sample can be subjected to azimuthal rotation of up to ±120° and tilt rotation from +30° to −15°, measured from the plane normal to the polar axis. To acquire the intrinsic information and clear dispersions of the electronic structure, most samples should be investigated at low temperature. The sample holder can be cooled using a continuous-flow cryostat, and it can be cooled to 7 K with a liquid helium cryostat.

A great challenge of small-beam-spot ARPES is that the exact position of the probing area might be lost during sample rotation. In fact, as the spot size and region of interest become smaller, this displacement becomes more significant in high-resolution ARPES experiments. To solve this problem, a high-performance hemispherical DA30 analyzer provided by Scienta Omicron GmbH is adopted; it has deflector plates in the lens system; thus, angles in a range of ±15° can be detected without rotating the sample.

In addition, a functional and user-friendly operation interface is a critical part of ARPES experiments and is essential for efficient data acquisition. A typical crystal sample measurement sequence is as follows. First, the sample is transferred to the pre-chamber and is cooled from room temperature to the target temperature to avoid contamination of the cleavage surface owing to gas absorption during the cooling process. When the temperature is stable, the crystal is cleaved to expose a flat, shiny surface, which can be checked by LEED if necessary. Most of this sample transfer process is automatically controlled by highly customized equipment, which can greatly reduce the workload of users as well as possible human errors. Users can easily move the manipulator to different positions by clicking buttons in the software interface. An interlock in the software has been set to avoid apparatus damage caused by misoperation.

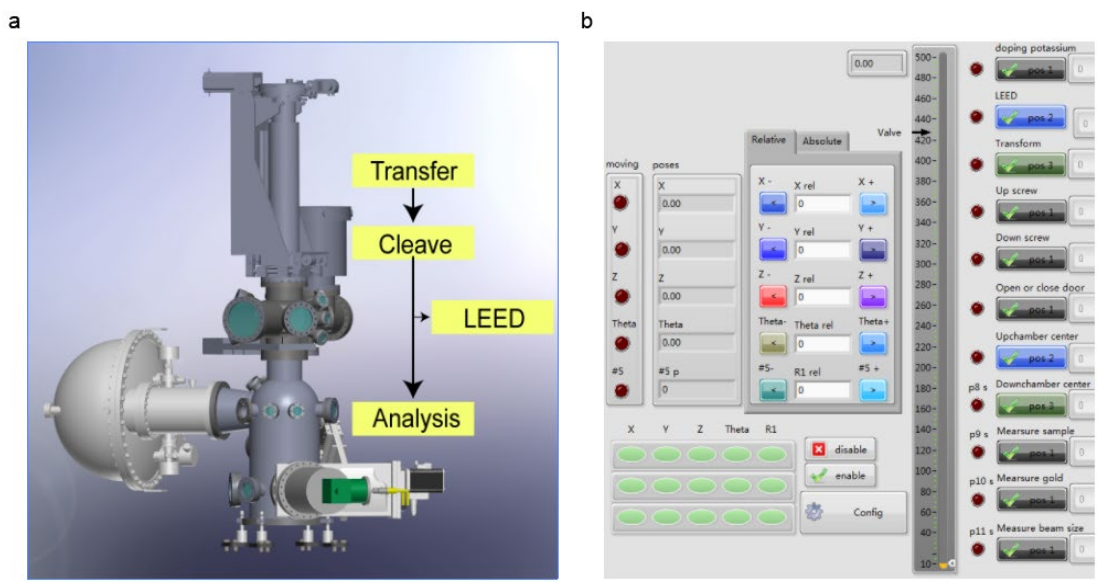

**Fig. 4. a** Sketch of ARPES system and typical measurement process. **b** Operation interface of the manipulator control software.

### 3.3 Oxide molecular beam epitaxy

A single-chamber OMBE system consisting of one UHV growth chamber and one ozone delivery system is now in use at our endstation. The growth chamber is equipped with a five-axis manipulator, which provides translational motion in three directions and rotational motion about two axes and enables easy sample transfer and reflection high-energy electron diffraction (RHEED) alignment; the sample holder can be heated to 900 ℃. The OMBE system shares a cassette load-lock chamber and UFO transfer chamber with the ARPES system, which enables quick loading of the substrate into the growth chamber under UHV conditions. The growth chamber is equipped with eight effusion cells and two single-pocket electron guns. An *in-situ* RHEED system is employed to monitor the growth of thin films. A quartz crystal microbalance is applied to measure the absolute deposition rates before and after film growth. The ozone delivery system integrated into the OMBE can supply ozone, which is more oxidizable than oxygen, for oxygenation during the process. Electro-pneumatic linear shutters provide accurate control of layer-by-layer deposition, with a precision of a small fraction of an atomic monolayer. Furthermore, the modular design of the OMBE system allows for easy and fast reloading or replacement of each source. This provides excellent flexibility for the deposition of different complex oxide compounds, including high-temperature superconductors. The base pressure of the OMBE chamber is $2 \times 10^{-10}$ mbar and rises to $2 \times 10^{-9}$ mbar when the source is heated to working temperatures.

The thin films grown in the OMBE chamber can be transferred through the UFO transfer chamber to the ARPES chamber instantly to realize *in-situ* ARPES measurement. Moreover, samples can also be treated by sputtering-annealing cycles in a minichamber by an ion source (SPECS IQE11/35).

## 3.4 Scanning tunneling microscopy/spectroscopy

The home-built scanning tunneling microscope system at BL03U provides a UHV and low-temperature working environment for studying the atomic and electronic structure of a conducting surface. It consists of the scanning tunneling microscope and three supporting subsystems: cryogenic, UHV, and vibration isolation systems. The microscope features the well-known Pan-type STM head design, which enables unprecedentedly high precision and repeatability in fine scanning and coarse approach. The spatial resolution thus achieved is subatomic, and the energy resolution can be as good as 0.1 meV. *In-situ* sample and tip exchange shorten the turnaround time, which is very important for a user facility. The cryogenic system includes a $^3$He refrigerator, a 14 T superconducting magnet, and a superinsulated liquid helium cryostat. The base temperature of 380 mK can be maintained for nearly 48 h. For delicate measurements that require even less vibration noise, the refrigerator can run in quiet mode with a base temperature of 440 mK for more than 12 h. The bipolar superconducting solenoid magnet provides a continuously tunable magnetic field (+14 to −14 T) normal to the sample surface, facilitating studies involving the spin or magnetism. The liquid-nitrogen-free cryostat avoids vibration from liquid nitrogen boil-off while maintaining low liquid helium consumption to maximize the working hours available without interruption due to liquid helium refilling. The UHV system generates and maintains a UHV ($2 \times 10^{-10}$ mbar) to protect the sample surface from contamination. It is equipped with standard sample/tip treatment accessories including an ion sputtering source, radiative heating (sample), and electron beam heating (tip). In addition, a spin-polarized STM tip and sample can be prepared by thermal evaporation with Knudsen cells. The entire system is connected to the UFO chamber and thus, eventually, to the OMBE and ARPES systems. High-quality thin films can be grown using the OMBE system and transferred to the STM and ARPES systems to observe the electronic behavior in real and momentum space, respectively. As another option, crystal samples can also be cleaved *in situ* in each system. To examine the STM spatial resolution, a NbSe$_2$ single crystal sample was measured at a temperature $T$ of 5 K with a bias voltage of 100 mV and constant current $I$ of 1 nA (Fig. 5). The vertical spatial resolution was 5 pm, and the transverse spatial resolution was 10 pm. In addition, a scanning tunneling spectrum (STS) was taken along the red line in Fig. 5a at $T = 440$ mK. The clear observation of the superconducting coherence peaks of NbSe$_2$ at ±0.6 meV demonstrates the excellent energy resolution of the STS.

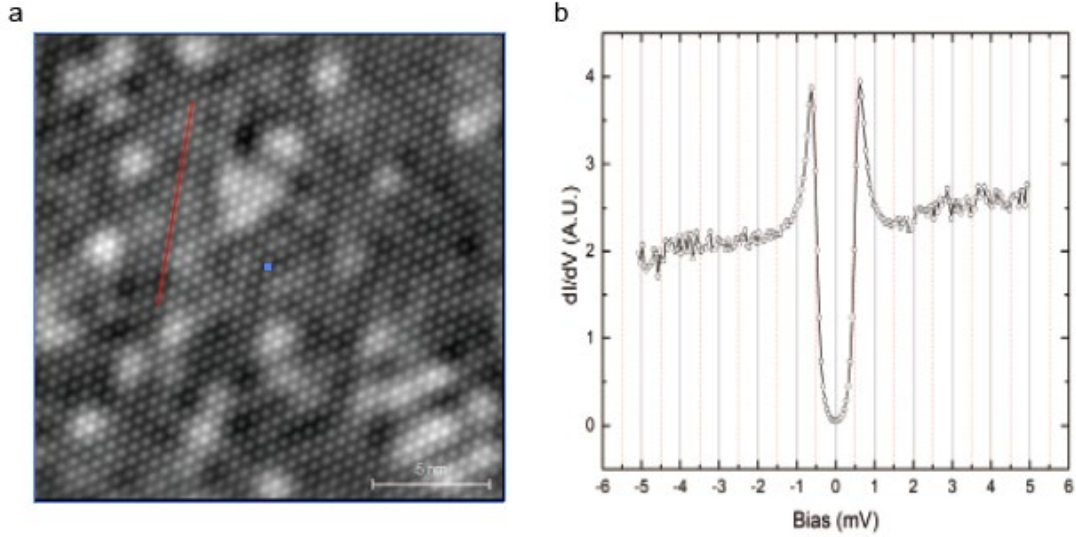

**Fig. 5. a** High-resolution STM image of NbSe$_2$ at $T$ = 5 K, $V$ = 100 mV, $I$ = 1 nA. **b** Experimental STS along the red line in **a** at $T$ = 440 mK.

## 4 Results and Discussion

### 4.1 ARPES performance

In a state-of-the-art ARPES system, the energy and angular resolution directly determine the performance of the equipment and are the most important parameters. Therefore, to obtain high resolution, many improvement measures have been adopted. The instrumental energy resolution is essentially determined by the linewidth of the light source and the low temperature of the sample. Hence, a thermal shield for the sample stage was developed, and the sample holder can be cooled to 7 K. In addition, there are two important requirements for obtaining the optimal performance of the electron analyzer: reliable electrical grounding and a stable power supply. The electrical grounding is quite essential for high energy resolution, as we have confirmed that the energy resolution can easily deteriorate to approximately 10 meV if the electrical grounding is not appropriate. This behavior probably stems from two interference factors. One is stray electrical noise, and the other is temporary charging of the sample owing to electron loss during the photoemission process. To avoid these effects, it was ensured that the cryostat, analyzer, and all chambers, especially the components with high voltage, are electrically connected to each other within a distance of 300 mm by a copper braid connected directly to ground. Furthermore, the analysis chamber has a double-deck μ-metal shield, which is contiguous with the μ-metal-shielded hemispherical electron energy analyzer to counteract the geomagnetic field. To test the energy resolution, we measured polycrystalline gold at $T$ = 7 K (Fig. 6a). The spectrum was analyzed by fitting it to a Fermi–Dirac occupation function convoluted with a Gaussian function. For a photon energy of 21.2 eV and a pass energy of 2 eV, we estimate the combined energy resolution to be 2.67 meV (with an analyzer energy resolution $\Delta E_{\text{ana}}$ of 1.12 meV). When the VUV laser is used, its energy resolution is better than that of synchrotron radiation, and it can be evaluated as 1 meV. In addition, the photoemission spectra of the typical topological insulator Bi$_2$Se$_3$ obtained using a 6.5 eV laser and 50 eV synchrotron radiation exhibit excellent momentum resolution (Fig. 6b,c). The acceptance angle and angular resolution of the analyzer were examined using a tungsten grating, and it was found that the maximum acceptable angle is ±15°, and the angular resolution is better than 0.1°.

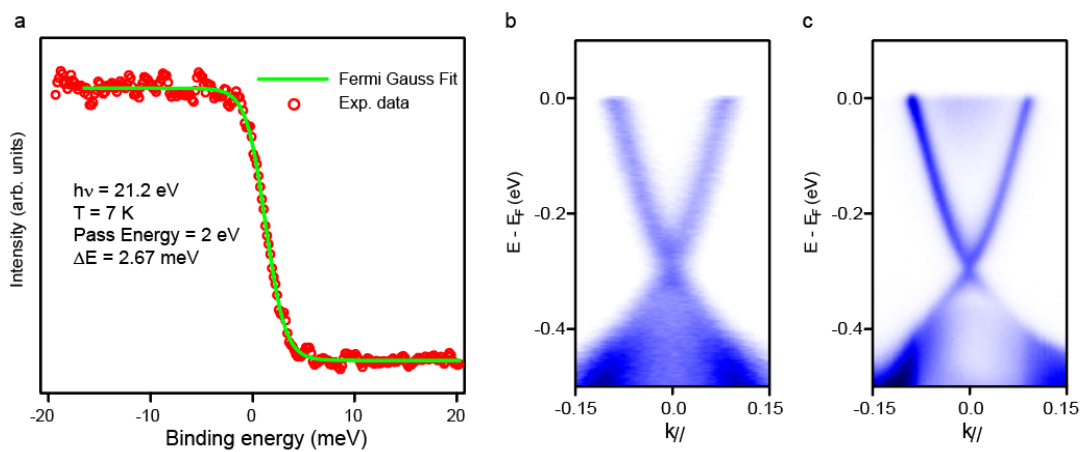

**Fig. 6. a** Measured photoemission spectrum of polycrystalline gold using a photon energy $h\nu$ of 21.2 eV. **b and c** ARPES data for topological insulator $Bi_2Se_3$ obtained using 50 eV synchrotron radiation and 6.5 eV VUV laser, respectively.

The spot size of the photon beam is also an important parameter of ARPES systems. A small beam spot can improve the momentum resolution, especially for multidomain samples. As mentioned in sec 2.1, we used a square aperture at the bottom of the manipulator to scan a photon beam in the vertical and horizontal directions and recorded the variation in photocurrent. Fig. 7 shows the results of first-order derivation of the experimental data and Gaussian fitting for a photon energy of 21.6 eV and silt size of 30 μm, which yield a spot size of 7.5 μm (V) × 67 μm (H) FWHM. This method is also used to measure the spot size of the VUV laser. The beam spot is circular, and the size is 30 μm (V) × 30 μm (H) FWHM.

In addition, as a useful supplement to synchrotron radiation, the VUV laser has other advantages. First, the high photon flux can cover at least $1 \times 10^{15}$ photons/s/0.1%B.W., which is as much as two orders of magnitude higher than that of synchrotron radiation. Furthermore, the escape depth of photoelectrons in the VUV region is longer than that of the synchrotron radiation of this beamline and is suitable for bulk-sensitive band measurements.

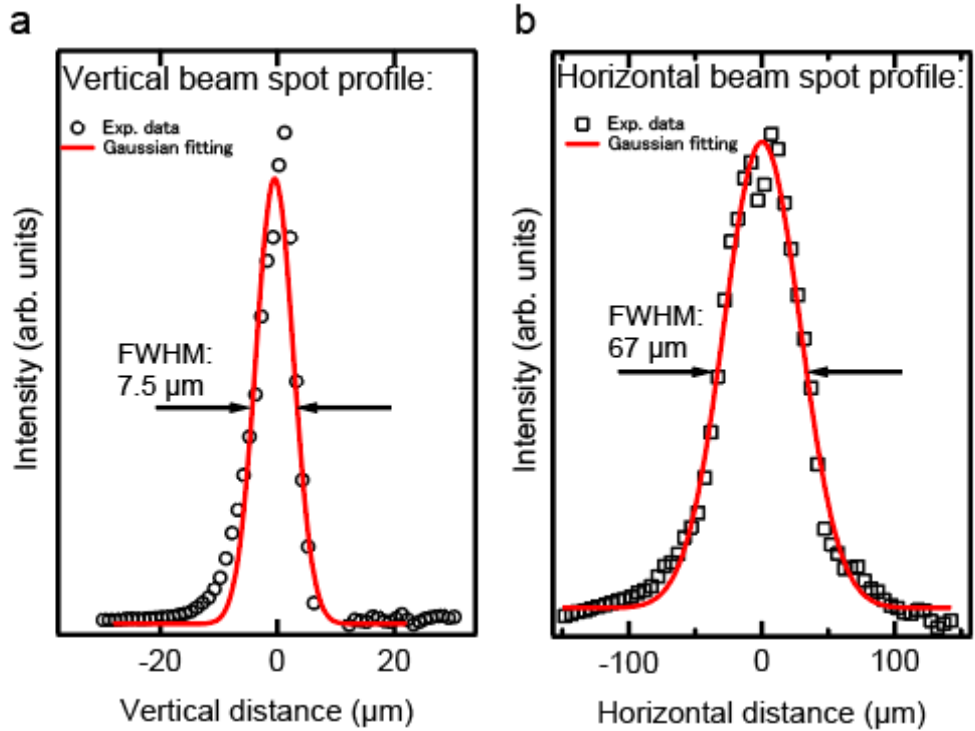

**Fig. 7.** Spot size of beam at sample position for photon energy of 21.2 eV. The FWHM is obtained from a Gaussian fitting. **a** Vertical beam spot for 7.5 μm. **b** Horizontal beam spot for 67 μm. Reprinted from **Ref. 24** with permission.

## 4.2 Applications to materials
### 4.2.1 *In-situ* electronic structure study of iridates

The first example illustrates the ability of the OMBE-ARPES system to realize thin film growth and *in-situ* electronic structure investigation of iridates. Among correlated electron systems, perovskite iridates, which host substantial electronic correlation and spin–orbit coupling, provide a platform for the study of the interplay between them [36–39]. Using the integrated OMBE+ARPES system, iridate thin films can be prepared under UHV conditions and transferred *in situ* to the analysis chamber for ARPES measurement. We applied OMBE to synthesize a series of high-quality orthorhombic perovskite iridates with Ruddlesden–Popper structure, $A_{n+1}Ir_nO_{3n+1}$, including $Sr_2IrO_4$ ($n$ = 1), $SrIrO_3$ ($n$ = ∞), and the superlattice $[(SrIrO_3)_m/SrTiO_3]$, $m$ = 1/2, 1, 2, and 3, and performed *in-situ* ARPES to study the low-lying electronic structure of the films [40]. Various characterizations, including RHEED, X-ray diffraction (XRD), and high-angle annular dark field scanning transmission electron microscopy (HAADF-STEM) analyses (Fig. 8a,b,c) show the high quality of the films. Z.-T. Liu et al. [41] grew $SrIrO_3$ thin films and used a He lamp to explore their electronic structure. The crossing of the electron band and hole band at the Fermi surface preliminarily confirmed the semimetallic state of $SrIrO_3$. Furthermore, the electron–plasmon interaction caused by photoelectron emission has also been investigated in perovskite $SrIrO_3$ [42]. By utilizing the *in-situ* OMBE-ARPES system, $SrIrO_3$ films of different thicknesses and different doping ratios were synthesized and studied by *in-situ* ARPES. A clear plasmonic polaron replica band due to electron–plasmon interaction is observed in the photoemission intensity plots.

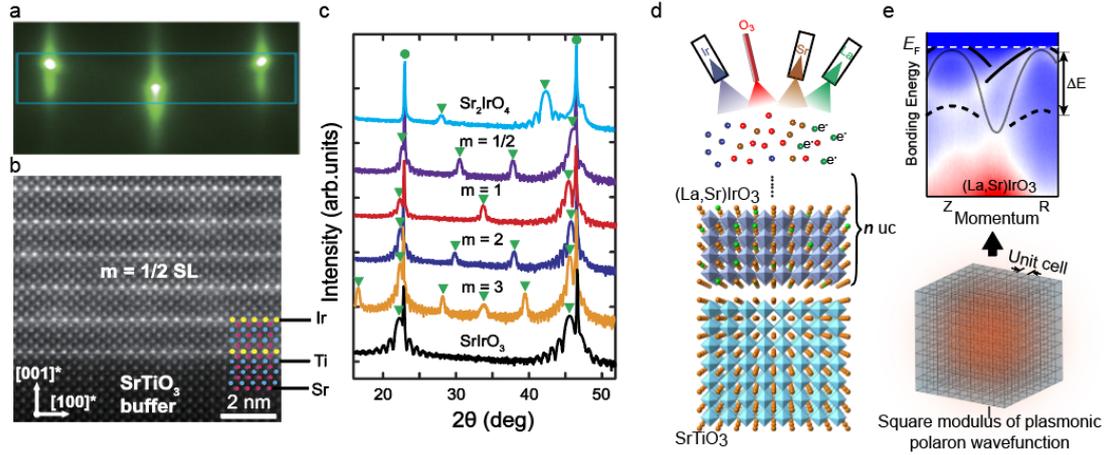

**Fig. 8. a** Typical RHEED patterns of 20 u.c. $m$ = 1 superlattice [(SrIrO$_3$)$_m$/(SrTiO$_3$)] taken along [100]$_p$ azimuth direction. **b** Atomically resolved HAADF-STEM image of $m$ = 1/2 superlattice thin films. **c** Typical XRD wide-range scan of Sr$_2$IrO$_4$ and $m$ = 1/2, 1, 2, 3 superlattice and SrIrO$_3$. **d** Schematic of fine carrier concentration tuning in La-doped SrIrO$_3$ thin films grown on SrTiO$_3$ substrate. **e** Band dispersion along Z-R high-symmetry direction. Solid and dashed lines represent main bands and their replicas, respectively. **a, b,** and **c** are reprinted from **Ref. 40** with permission. **d** and **e** are reprinted from **Ref. 42** with permission.

**4.2.2 Fine electronic structure study of magnetic topological insulator MnBi$_2$Te$_4$**

The second example demonstrates the ability of ARPES in the VUV region to realize investigations of the fine electronic structure of the intrinsic magnetic topological insulator MnBi$_2$Te$_4$. As a frontier field in condensed matter physics, the combination of topology and magnetism is predicted to reveal various novel quantum states, such as the quantum anomalous Hall effect and topological axion insulating states [43–45]. The recently discovered MnBi$_2$Te$_4$ family, which has layered crystalline structure and antiferromagnetic order, is a promising intrinsic magnetic topological insulator candidate [46]. Thus, direct observation of the fine electronic structure of the MnBi$_2$Te$_4$ family, especially the Dirac surface states, is necessary to obtain a comprehensive picture of the novel quantum states resulting from the combination of magnetic order and topology.

Three groups have performed synchrotron ARPES measurements of MnBi$_2$Te$_4$ at this BL03U beamline. Li et al. [23] used 13.8 eV photons to measure the electronic structure of MnBi$_2$Te$_4$ at different temperatures. Both the ARPES data and the second derivative reveal a linear crossing of two bands that form a Dirac point at −0.28 eV (Fig. 9a–9d). Interestingly, the electronic structure of the paramagnetic and antiferromagnetic phases of MnBi$_2$Te$_4$ is the same (Neel temperature $T_N$ = 24 K), and they suggest that the reason is weak coupling between the local magnetic moments and the topological electronic states. Y.-J. Chen et al. [24] also observed gapped bulk electronic bands and topological surface states with a diminished gap. B. Chen et al. [25] used high-resolution ARPES at BL03U to investigate the low-energy electronic structure of antimony-doped Mn(Sb$_x$Bi$_{1−x}$)$_2$Te$_4$ samples. Fig. 9e shows the ARPES intensity plot around the center of the Γ point of the Brillouin zone at a photon energy of 7.25 eV. In this plot, the parabolic dispersions and Dirac-cone-like topological surface state can be clearly identified. Notably, the constant-energy counters display the evolution of the surface state, and the Dirac-like point is located at an energy 260 meV below the Fermi level (Fig. 9g). In addition, samples with different antimony ratios were measured with 14 eV photons. At *x* = 0, the band gap is below the Fermi level, indicating heavy n-type doping. When *x* = 0.3, the Fermi surface crosses the band gap, and the carriers are transformed from n-type to p-type by the square (Fig. 9h–9k). A phase diagram showing the properties of materials with different doping ratios was constructed using experimental data. It revealed that there is a range in which perfect topologically nontriviality, initial magnetic order, and bulk-phase carrier suppression are obtained. These achievements provide important insights for further research on magnetic topological insulators.

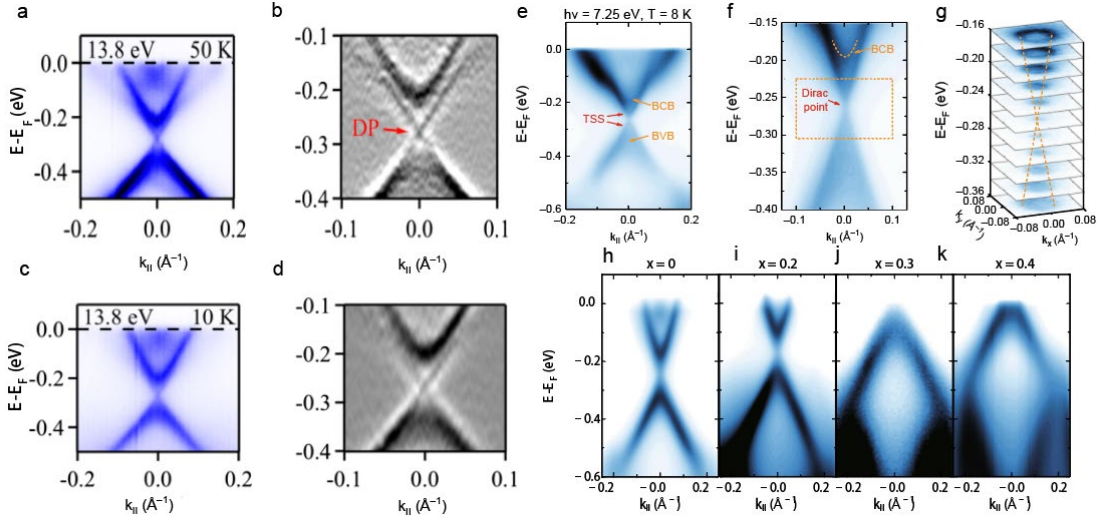

**Fig. 9. a,c** ARPES intensity maps along the cut through Γ taken at *hν* = 13.8 eV at 50 and 10 K. **b,d** Intensity maps of the second derivative with respect to the energy of the data near the bulk band gap in **a** and **c**. DP: Dirac point. **e,f** ARPES data of the crystal and the corresponding magnified view near the Dirac point at a photon energy of 7.25 eV and 8 K. **g.** Constant-energy maps at binding energies of −0.16 to −0.36 eV. **h–k.** ARPES data of samples with different antimony substitution ratios. **a–d** are reprinted from **Ref. 23** with permission, and **e–k** are reprinted from **Ref. 25** with permission.

### 4.2.3 Electronic structure study with surface adsorption of iron-based superconductor CaKFe$_4$As$_4$

The third example illustrates band tuning by surface doping, which is used to investigate the iron-based superconductor CaKFe$_4$As$_4$. Iron-based superconductors are attracting intense interest owing to their high superconducting transition temperature $T_c$ and wide range of compositional tunability [47–52]. In the iron-based superconductor family, iron pnictides are predicted to have a higher superconducting $T_c$ and more abundant crystal structure and homogeneity [53]; thus, they are considered to be promising candidates for Majorana zero modes.

Recently, W.-Y. Liu *et al.* [21] performed a comprehensive study of a new iron pnictide superconductor, CaKFe$_4$As$_4$ by ARPES with *in-situ* potassium surface adsorption. They used the high-resolution ARPES system at BL03U with a photon energy of 76 eV, which has a small mean free path of photon-emitted electrons, to investigate the superconducting Dirac surface states. Fig. 10b shows the experimental ARPES image of undoped CaKFe$_4$As$_4$, where the selected $k_z$ is around the Γ point. The ARPES data show that the 3d$_z^2$/4p$_z$ band (blue line) is far below the Fermi level, and an extra band (in the red box) is distinguished from it. The corresponding second-order derivative plot also reveals this band, which has a linear-like dispersion, implying that this band is contributed by surface states. To verify the presence of a topological surface state, they utilized potassium surface adsorption on the sample for 40 s, which can afford charge doping to raise the Fermi level. Notably, the surface-state dispersion is clearer in the K-doped sample than the undoped sample (Fig. 10c), and the Dirac-cone-like band can be observed in both the second-order derivative plot and the momentum distribution curve (Fig. 10e,f). The result of the K-doping experiment confirms the presence of a topological surface state, which may induce a topological vortex with Majorana zero modes.

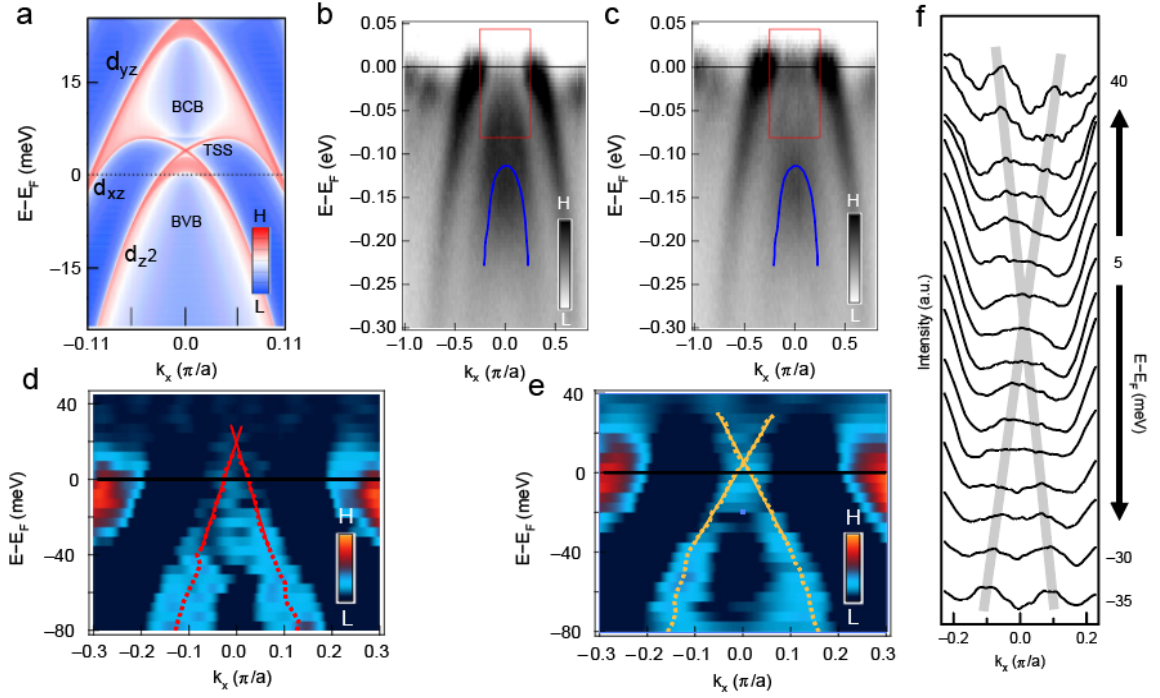

**Fig. 10. a** Calculated band structure projected onto the (001) surface. **b** ARPES spectral intensity plot along the Γ−M direction of undoped sample with a photon energy of 76 eV at BL03U at a temperature of 25 K. **c** Same as **b**, but the surface of the sample is K-doped. **d, e** MDC second-order derivative plot of **b,c** in the red box. **f** MDC plot of **c**, where gray line are guides to indicate the positions of peaks. Reprinted from **Ref. 21** with permission.

## 4 Conclusion

Several ARPES beamlines and endstations have been constructed at third-generation synchrotron radiation facilities, such as X03MA in Swiss Light Source [9] and I05 in Diamond Light Source [10], and have contributed many scientific advances. With the development of advanced detectors and progressive light sources, we can expect the exploration of higher energy and momentum resolutions, small spot sizes, and new degrees of freedom by ARPES. Spin-resolved detectors and time-of-flight spectrometers will propel ARPES to a new stage [54–56]. In addition, the combination of sample growth techniques with ARPES provides more possibilities for investigation. Using the OMBE growth method, one can use numerous methods to tune parameters, such as the strain from the substrate, the elementary doping and thickness of thin films, and artificial heterostructures, which provide infinite opportunities for future band engineers.

To summary, we constructed a new ARPES beamline and an endstation BL03U at the SSRF and briefly introduce several recent scientific studies. The system utilizes a synchrotron radiation light sources and VUV laser, which is equipped with a DA30 analyzer. At low energy, our ARPES system can realize best energy resolution at 2.67 meV and a small beam spot of 7.5 μm (V) × 67 μm (H). The very small beam spot provides a promising opportunity for advanced ARPES measurement. In addition, the integration of STM and OMBE can support multidimensional investigations of the electronic structure of solids, including topological quantum materials, *in-situ*-grown oxide thin films, superconducting and exfoliated two-dimensional materials. Our endstation is part of the project SiP.ME$^2$, which can provide more *in-situ* measurement techniques (X-ray photoelectron, absorption, and emission spectroscopy) [57,58]. The construction of this project is complete, and it has been open to users since April 2019.

**Contribution statements**

All authors contributed to the study conception and design. The endstation design and construction were performed by Zheng-Tai Liu, Zhong-hao Liu, Ji-Shan Liu, Wan-Ling Liu, Xiang-Le Lu and Da-Wei Shen. The beamline was designed and constructed by Mao Ye and Qiao Shan. The STM system was designed and constructed by Hong-Ping Mei and Ang Li. The first draft of the manuscript was written by Yi-Chen Yang, and all authors commented on previous versions of the manuscript. All authors read and approved the final manuscript.

**Acknowledgments:** We gratefully acknowledge financial support from the National Key R&D Program of the MOST of China (Grant No. 2016YFA0300204) and the National Natural Science Foundation of China (No. 11227902) as part of the SiP·ME2 beamline project.